\documentstyle[12pt,fleqn,epsfig]{article}
\begin{document}

\begin{center}
{\bf INSTITUT~F\"{U}R~KERNPHYSIK,~UNIVERSIT\"{A}T~FRANKFURT}\\
D - 60486 Frankfurt, August--Euler--Strasse 6, Germany
\end{center}

\hfill IKF--HENPG/2--99


\vspace{1.5cm}
\begin{center}
   {\Large \bf Evidence for  Statistical Production of $J/\psi$ Mesons   
   in Nuclear Collisions at the CERN SPS  }  
\end{center}

\vspace{1cm}

\begin{center}

Marek Ga\'zdzicki\footnote{E--mail: marek@ikf.physik.uni--frankfurt.de}\\
Institut f\"ur Kernphysik, Universit\"at Frankfurt,
Germany\\[0.8cm]

Mark I. Gorenstein\footnote{Permanent address: Bogolyubov Institute for 
Theoretical Physics,
Kiev, Ukraine}$^,$\footnote{E--mail: goren@th.physik.uni-frankfurt.de}\\
Institut f\"ur Theoretische Physik, Universit\"at  Frankfurt,
Germany

\end{center}

\vspace{1cm}

\begin{abstract}
\noindent
The hypothesis of  statistical production of
$J/\psi$ mesons at hadronization is formulated and
checked against experimental data.
It explains in the natural way the observed
scaling behavior of the $J/\psi$ to pion ratio
at the CERN SPS energies.
Using the multiplicities of $J/\psi$ and $\eta$ mesons
the hadronization temperature $T_H \cong 170$ MeV
is found,
which agrees well with the previous estimates 
of the temperature parameter based on the analysis of
the hadron  yield systematics.
\end{abstract}

\newpage

Charmonium production in hadronic 
\cite{Ma:95} and nuclear \cite{Vo:99} collisions
is usually considered to be  composed of three  stages:
the creation of a $c\overline{c}$ pair, the formation of a
bound $c\overline{c}$ state and the subsequent interaction 
of this  $c\overline{c}$ bound state with the
surrounding matter. 
The first process is calculated within perturbative QCD,
whereas modeling of  non--perturbative dynamics is needed to describe 
the last two stages (see, e.g. \cite{Ge:98} and references therein).
The interaction of the bound $c\overline{c}$ state with  matter
causes suppression of the finally observed charmonium yield
relative to the initially created number of bound $c\overline{c}$
states. This initial number is assumed to be proportional to the number
of Drell--Yan pairs, which then allows for the experimental study
of the charmonium suppression pattern.
It was proposed
\cite{Sa:86,Sh:78} that the magnitude of the measured suppression
in nuclear collisions can be used as a probe of the state of
high density matter created at the early stage of the collision.
The suppression of the $J/\psi$ yield  observed in p+A and O(S)+A collisions
at the CERN
SPS is considered to be caused by the interactions with 
nucleons  occurring while the primordial baryons keep
interpenetrating \cite{Ge:92}. 
The rapid increase of the suppression
({\it anomalous suppression}) observed when going from peripheral
to central Pb+Pb collisions \cite{NA50} 
is often attributed to the formation
of a quark--gluon plasma 
\cite{Sa:97}. 
However alternative interpretations 
are still  under discussion
\cite{Vo:99,Ge:98,Sp:99}. 

\vspace{0.2cm}

\begin{figure}[p]
\epsfig{file=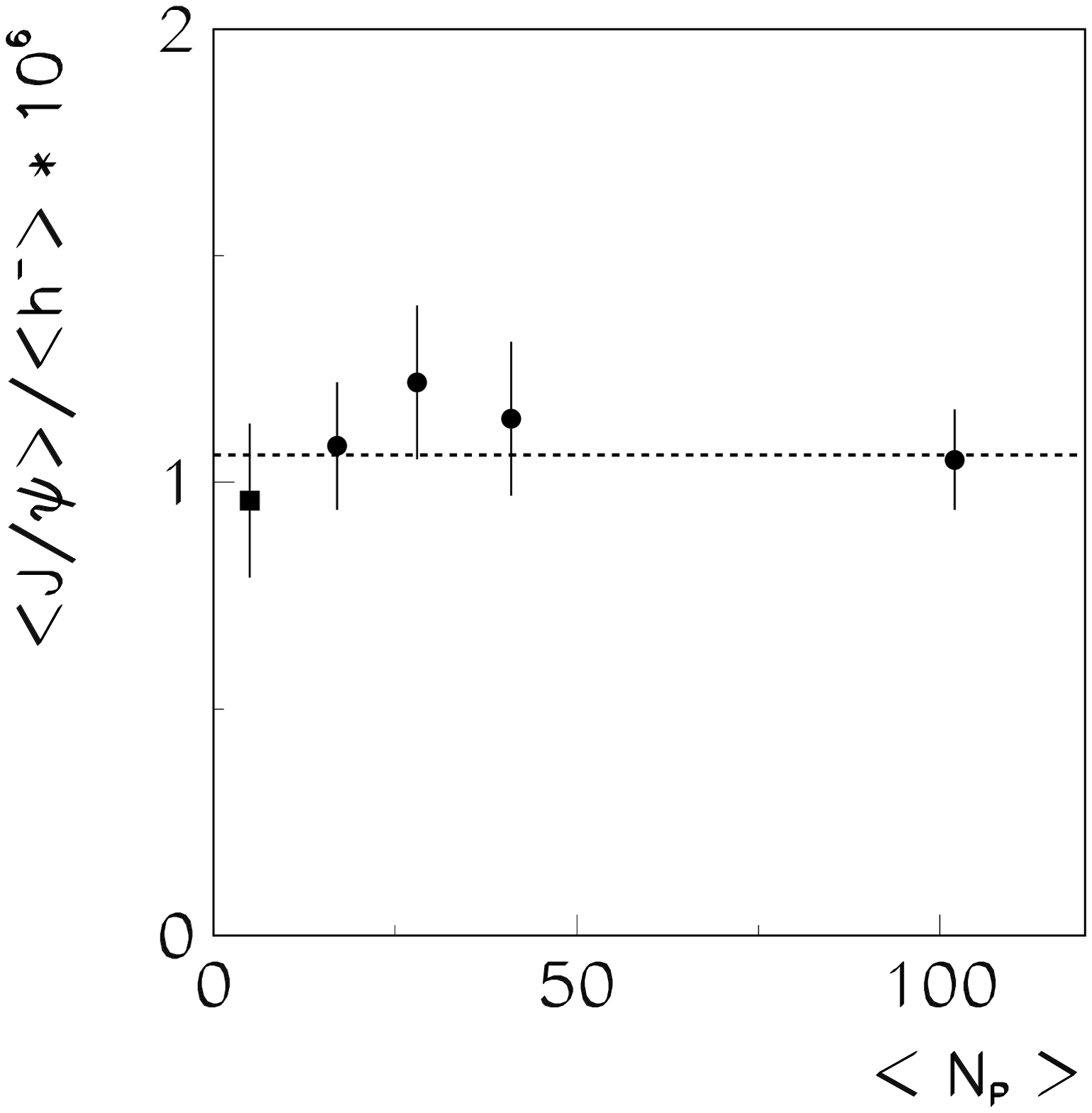,width=14cm}
\caption{
The ratio of the mean multiplicities of $J/\psi$ mesons
and negatively charged hadrons for inelastic nucleon--nucleon (square) and
inelastic O+Cu, O+U, S+U and Pb+Pb (circles) interactions at   
158 A$\cdot$GeV plotted as a function of the mean
number of participant nucleons.
For clarity the N+N point is shifted from
$\langle N_P \rangle = 2$ to $\langle N_P \rangle = 5$.
The dashed line indicates the mean value of the ratio.
}
\label{fig1}
\end{figure}

It was recently found \cite{Ga1:98,Ga2:98} 
that the mean 
multiplicity of $J/\psi$ mesons increases proportionally to the mean
multiplicity of pions when proton--proton (p+p), 
proton--nucleus (p+A)
and nucleus--nucleus (A+A) collisions at CERN SPS energies are 
considered.
We illustrate this unexpected experimental fact by reproducing in
Fig. \ref{fig1} the plot from Ref. \cite{Ga2:98}, where
the ratio $\langle J/\psi \rangle / \langle h^- \rangle$ is
shown as a function of the mean number of nucleons participating
in the interaction for inelastic nuclear collisions at the
CERN SPS. 
The $\langle J/\psi \rangle$ and  $\langle h^- \rangle$ denote
here the
mean multiplicites of $J/\psi$ mesons and negatively
charged hadrons (more than 90\% are $\pi^-$ mesons), respectively.
We note that the analysis presented in Ref. \cite{Ga1:98}
indicates that the scaling of the 
$\langle J/\psi \rangle/\langle h^- \rangle$ ratio is also valid 
for central Pb+Pb collisions at the CERN SPS.

\vspace{0.1cm}
\noindent
In the standard picture of the $J/\psi$ production
based on  the {\it hard creation } of $c\overline{c}$ pairs  
and the 
subsequent 
{\it  suppression } of the bound $c\overline{c}$  states 
the observed scaling behavior of the  $J/\psi$
multiplicity  
appears to be  due to 
an `accidental'  cancelation of  several large effects.
This motivates our effort to find an alternative production 
mechanism of $J/\psi$ mesons which would explain
the experimental data  in a natural way.

\vspace{0.1cm}
\noindent
In this letter we show that a scaling property of the
$J/\psi$ multiplicity 
\begin{equation}\label{scaling}
\frac {\langle J/\psi \rangle}  {\langle h^- \rangle}~\cong~ 
const(A)
\end{equation}
can be understood assuming that a dominant fraction of 
$J/\psi$ mesons is produced directly at hadronization
according to the available hadronic phase space.

\vspace{0.2cm}

Since a long time \cite{Fe:50} statistical models
are used to describe hadron multiplicities
in high energy collisions.
Thermal hadron production models have been successfully  
used to fit the data on particle multiplicities
in A+A collisions at the CERN SPS energies (see, e.g. 
\cite{Go:99,Be:98}).
Due to the large 
number of particles a grand canonical formulation
is used for the modeling of high energy heavy 
ion collisions \cite{Ra:80,Be:98}. 
Recently, an impressive success of the statistical model
applied to hadron multiplicities in elementary
$e^+ + e^-$, $p+p$ and $p+\bar{p}$ interactions at
high energy was also reported \cite{Be:96}.
However,
in the latter case the  use of a canonical formulation of the
model, which assures exact conservation of the
conserved charges, is necessary.
The temperature parameter which characterizes
the available phase space  
for the hadron production 
is found in these interactions to be 160--190 MeV \cite{Be:96}.
It does not show any significant dependence on the type
of reaction and on the collision energy. 
Moreover,
it coincides with  the chemical freeze--out 
temperature estimate
obtained in hadron gas models for A+A collisions at the CERN 
SPS \cite{Go:99}. These facts suggest the  possibility to ascribe
the observed statistical properties
of hadron production systematics in elementary and nuclear collisions
at high energies to  the statistical nature
of the hadronization process 
\cite{Be:96,Be:98,St:99}.

\vspace{0.2cm}

Based on the above facts we formulate
a hypothesis that
{\bf a dominant fraction of the $J/\psi$ mesons produced in hadronic and
nuclear collisions at the CERN SPS energies is created 
at hadronization according to the available hadronic phase space.}

\vspace{0.1cm}
\noindent
$J/\psi$ mesons are neutral and unflavored, i.e. all charges 
conserved in the strong interaction  
(electric charge, baryon number, strangeness and charm)
are equal to zero for this particle.
Therefore, its production is not influenced by the conservation laws of
quantum numbers.
For suffiently high collision energies, the effect of
the strict energy--momentum conservation in the statistical model
formulation can be 
neglected. 
Consequently, the $J/\psi$ production can be calculated
in the grand canonical approximation and, therefore, its multiplicity is
proportional
to the volume, $V$, of the matter at hadronization. 
Thus, the statistical yield 
of $J/\psi$ mesons 
at hadronization 
is  given by
\begin{eqnarray}\label{stat}
\langle J/\psi \rangle~&=&~\frac{(2j+1)~V}{2\pi^2} \cdot
\int_0^{\infty}p^2dp~\frac{1}{\exp[(p^2+m_{\psi}^2)^{1/2}/T_H]~-~1}~\\
&\cong& ~ 
(2j+1)\cdot V \cdot 
\left(\frac{m_{\psi}T_H}{2\pi}\right)^{3/2} \cdot \exp\left(-~ 
\frac{m_{\psi}}{T_H}\right)~, \nonumber
\end{eqnarray}
where $j=1$ and $m_{\psi} \cong 3097$~MeV are the spin and the mass 
of the $J/\psi$ meson and
$T_H$ is the hadronization 
temperature.  
The previously mentioned  results of the analysis
of hadron yield systematics  in elementary and nuclear collisions 
within the statistical approach
indicate that the hadronization temperature $T_H$ 
is the same for different
colliding systems and collision energies. This reflects
the universal
feature of the hadronization process.\\
The total entropy of the produced matter is proportional to its 
volume. As most of the entropy in the final state
is carried by pions, the pion multiplicity is also expected to be 
proportional to the volume of the hadronizing matter.
Thus the scaling property (\ref{scaling}) follows directly 
from the hypothesis of  statistical production of $J/\psi$  mesons
at hadronization and the universality of the parameter $T_H$.\\
Since elements of hadronizing matter move in the overall center
of mass system
the volume $V$ in Eq.~(\ref{stat})
characterizes in fact the sum of the proper volumes of all
elements in the collision event. 

\vspace{0.2cm}

The hypothesis of  statistical production of $J/\psi$ mesons
at a constant hadronization temperature $T_H$ leads  
to the prediction of a 
second scaling property of the $J/\psi$ multiplicity, namely:
\begin{equation}\label{scaling2}
\frac {\langle J/\psi \rangle}  {\langle h^- \rangle}~\cong~ 
const(\sqrt{s})
\end{equation}
which should be valid for sufficiently large c.m.
energies, $\sqrt{s}$.
This scaling property is illustrated  in Fig. 2
which shows the ratio $\langle J/\psi \rangle/\langle h^- \rangle$
as a function of $\sqrt{s}$ for proton--nucleon interactions.
The experimental data on $J/\psi$ yields are taken from a compilation 
given in \cite{Sch:94}. The values of $\langle h^- \rangle$
are calculated using a parameterization of the experimental 
results as proposed in \cite{Ga:91}.\\
Onwards from the CERN SPS energies,
$\sqrt{s} ~\cong~ 20$ GeV, the ratio 
$\langle J/\psi \rangle/\langle h^- \rangle$ 
is approximately constant, in line with the expected scaling
behavior
(\ref{scaling2}).
The rapid increase of the ratio with collision energy
observed below $\sqrt{s} ~\cong~ 20$ GeV should be attributed
to a significantly larger energy threshold for the $J/\psi$
production than for the  pion production.
In terms of the statistical approach
the effect of strict energy--momentum conservation has to be
taken into account by use of the microcanonical formulation
of the model.

\vspace{0.2cm}

The statistical $J/\psi$ multiplicity (\ref{stat})
depends on two parameters, $T_H$ and $V$. 
In general the calculation of the hadron yields in the
statistical model should take into account  
the conservation of charges and resonance feeddown contributions.
However,
a simple way to estimate of the crucial temperature
parameter in Eq.~(\ref{stat}) 
from the experimental data is possible, provided that
we find a second hadron  which 
has the properties of the $J/\psi$ meson i.e.
it is neutral, unflavored and stable with respect to strong decays.
The best candidate is the $\eta$ meson.
The multiplicity of $\eta$ mesons seems to obey also the scaling
properties (\ref{scaling}) and (\ref{scaling2}). 
We note, however,
that the data on $\eta$ production are scarce.
The independence of the $\langle \eta \rangle/\langle \pi^0 \rangle$
ratio on the collision energy  was observed 
quite a long time ago \cite{Do:78}.
Recent data on $\eta$ production in central Pb+Pb collisions
at the CERN SPS \cite{Pe:98} suggest that the 
$\langle \eta \rangle/\langle \pi^0 \rangle$ ratio is also
independent of the size of the colliding objects.
In order to illustrate this scaling  Fig. 3 shows
the $\langle \eta \rangle/\langle \pi^0 \rangle$ ratio
as a function of the number of interacting nucleons
for inelastic p+p \cite{Ag:91}
and S+S \cite{Al:95} interactions and for central Pb+Pb \cite{Pe:98}
collisions at the CERN SPS energies (158--400 A GeV).

\vspace{0.1cm}
\noindent
From the ratios, 
$\langle J/\psi \rangle/\langle h^- \rangle$ and
$\langle \eta \rangle/\langle \pi^0 \rangle$, 
presented in Fig. 1 and Fig. 3 we estimate a mean
ratio $\langle J/\psi \rangle/\langle \eta \rangle =
(1.3 \pm 0.3) \cdot 10^{-5} $.
Here we use the experimental ratio 
$\langle \pi^0 \rangle/\langle h^- \rangle ~\cong~ 1 $ in
N+N interactions \cite{Ha:91}. 
Under the hypothesis of the statistical production of
$J/\psi$ and $\eta$ mesons at hadronization
the measured ratio can be compared to the ratio
calculated using Eq.~(\ref{stat}):
\begin{equation}\label{eta}
\frac{\langle J/\psi \rangle}{\langle \eta \rangle}~\cong~
 3 \cdot \left(\frac{m_{\psi}}{m_{\eta}}\right)^{3/2}
\cdot \exp\left(\frac{m_{\eta}-m_{\psi}}{T_H}\right)~,
\end{equation}
where $m_{\eta} \cong 547$~MeV is the mass of the $\eta$ meson.
This leads to an estimate of the hadronization temperature,
$T_H = 170 \pm 2$ MeV.
Assuming a maximum 50\% uncertainty on the 
$\langle J/\psi \rangle/\langle \eta \rangle$
ratio due to the contribution from  resonance decays\footnote{
An estimate of the fraction of the $\eta$ yield from
decays of heavy hadrons in p+p interactions is about 50\% \cite{Be:96},
which is close to the measured fraction of $J/\psi$ yield 
originating from  decays (30--50\%) \cite{Ab:97,Vo:99}.}
we obtain an estimate of an additional 
systematic error on $T_H$ of about 7 MeV. 
A graphical solution of Eq. (\ref{eta}) is shown in Fig. 4
which illustrates the high sensitivity of the estimate
of the $T_H$ parameter by using  the 
$\langle J/\psi \rangle/\langle \eta \rangle$ ratio.
This is due to the large difference between mass of the $J/\psi$
and the $\eta$ mesons.

\vspace{0.2cm}

In summary, we show that the $J/\psi$ production in hadronic
and nuclear collisions can be understood assuming that
a dominant fraction of $J/\psi$ mesons is produced at hadronization
according to the available hadronic phase space.
The estimate of the hadronization temperature based on 
$J/\psi$ multiplicity, $T_H \approx 170$ MeV, agrees well
with the values of the temperature parameter obtained
from the analysis of the hadron yield systematics
in 
$e^+ + e^-$, $p+p$, $p+\bar{p}$ interactions and
nucleus--nucleus collisions.

\vspace{0.2cm}

If the new interpretation of the $J/\psi$ data presented
in this letter is correct,  
it may have several important implications:

{\bf 1.} $J/\psi$ yields are not sensitive to the state of high
density matter created at the early stage of A+A collisions
because their production takes place at  hadronization.

{\bf 2.} The creation of the $J/\psi$ mesons is due to
the  straight thermal production at hadronization and not due to
the coalescence of $c \overline{c}$ quarks produced before 
hadronization. 
Therefore the
yield of $J/\psi$ mesons is independent 
of the production of open charm, which 
is carried
mainly by the $D$ mesons 
in the final state.
The $D$ meson multiplicity is determined
by the number of $c\overline{c}$ quark pairs created in the 
early {\it parton stage} before the hadronization. 

{\bf 3.} Due to the large mass of $J/\psi$ mesons the data
on their production are very sensitive to the value of 
the hadronization temperature and  
therefore allow for a precise study of the hadronization process.

\vspace{1cm}
\noindent
{\bf Acknowledgements}

We thank
F. Becattini,
L. Gerland, I. Mishustin, St. Mr\'owczy\'nski,
P. Seyboth, R. Stock, H. St\"ocker and G. Yen for 
discussion and comments to the manuscript.
We acknowledge
 financial support of BMBF and  DFG, Germany.

\newpage

\begin{figure}[p]
\epsfig{file=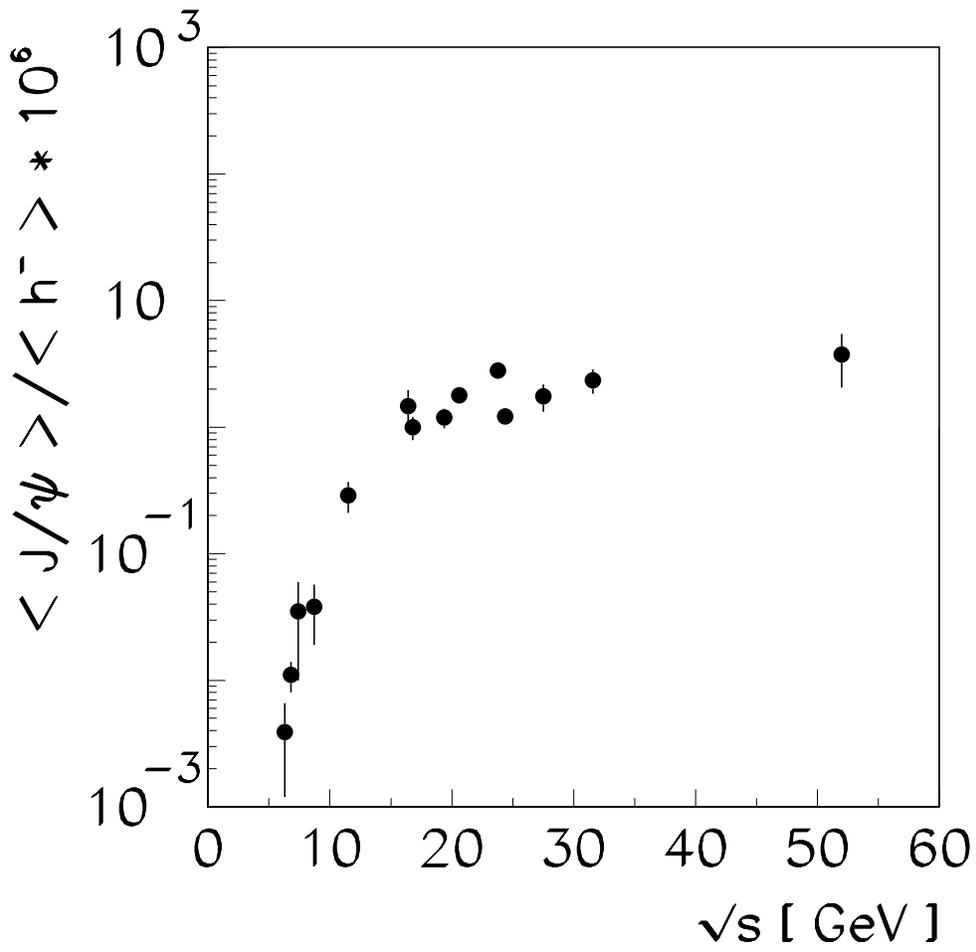,width=14cm}
\caption{
The ratio of the mean multiplicities of $J/\psi$ mesons
and negatively charged hadrons for inelastic proton--nucleon 
interactions  as a function of the collision energy in the
center of mass system.  
}
\label{fig2}
\end{figure}

\begin{figure}[p]
\epsfig{file=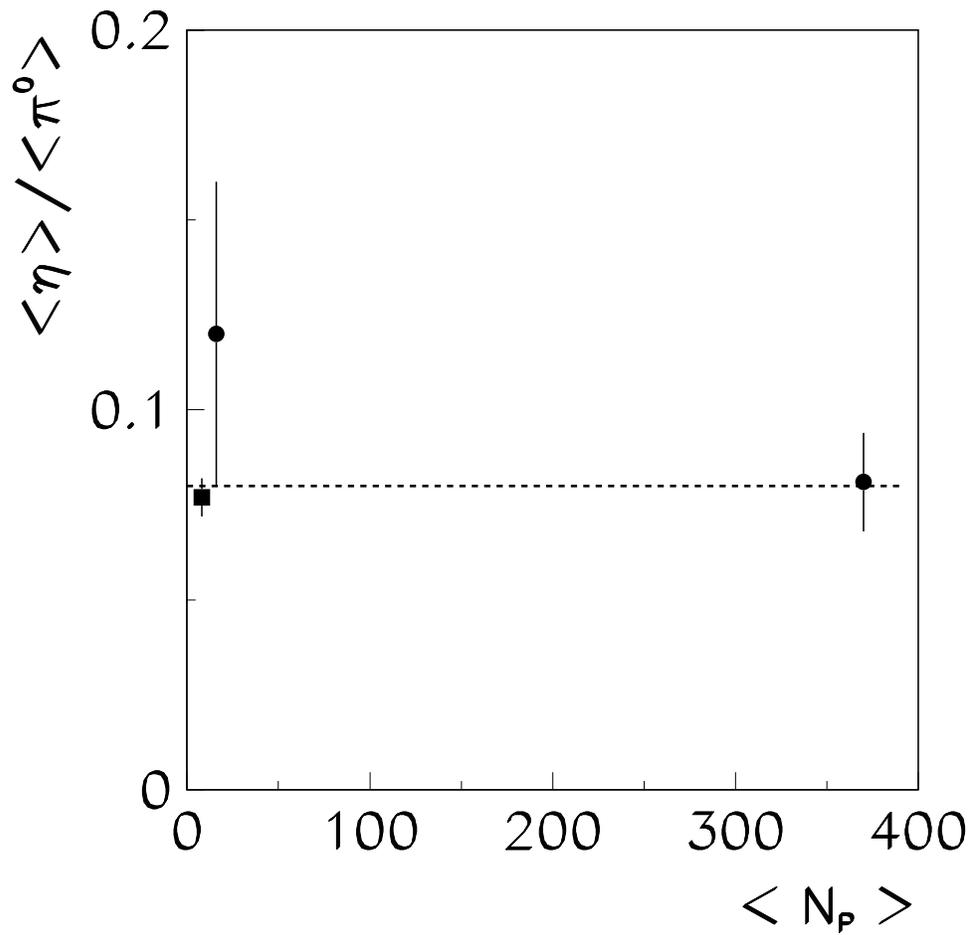,width=14cm}
\caption{
The ratio of the mean multiplicities of $\eta$ mesons
and $\pi^0$ mesons for inelastic p+p and S+S
interactions and central Pb+Pb collisions at the CERN SPS
energies as a function of the number of interacting nucleons. 
The result for Pb+Pb collisions was obtained in the
rapidity interval 2.3--2.9.
}
\label{fig3}
\end{figure}

\begin{figure}[p]
\epsfig{file=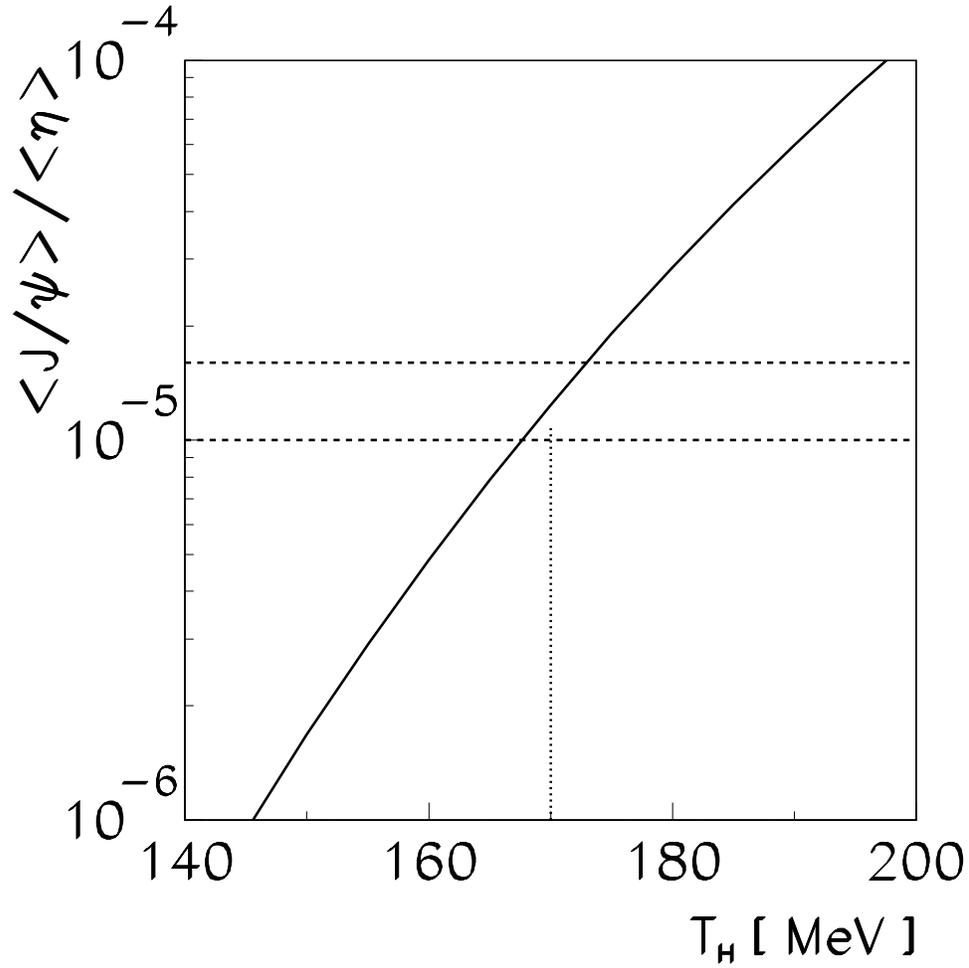,width=14cm}
\caption{
The $\langle J/\psi \rangle/\langle \eta \rangle$ ratio
calculated under hypothesis of the statistical production of
$J/\psi$ and $\eta$ mesons at hadronization (solid line)
as a function of the hadronization temperature.
Band shown by dashed lines is drawn at $\pm \sigma$
around the mean experimental value of the 
$\langle J/\psi \rangle/\langle \eta \rangle$ ratio.  
The dotted line indicates $T_H = 170$ MeV.
}
\label{fig4}
\end{figure}

\end{document}